\documentclass[%
 reprint,
showpacs,preprintnumbers,
 amsmath,amssymb,
 aps,
]{revtex4-1}

\usepackage{graphicx}
\usepackage{dcolumn}
\usepackage{bm}

\newcommand\redout{\bgroup\markoverwith{\textcolor{red}{\rule[.5ex]{2pt}{0.4pt}}}\ULon}

\usepackage[colorlinks=true,citecolor=blue]{hyperref}
\hypersetup{colorlinks=true,citecolor=blue,linkcolor=red,urlcolor=blue}
\begin{document}
\title{Topological phase and edge states dependence of the RKKY interaction in zigzag silicene nanoribbon}
\date{\today}

\author{Moslem Zare}
\affiliation{School of Physics, Institute for Research in Fundamental Sciences (IPM), Tehran 19395-5531, Iran}

\author{Fariborz Parhizgar}\email{fariborz.parhizgar@ipm.ir}
\affiliation{School of Physics, Institute for Research in Fundamental Sciences (IPM), Tehran 19395-5531, Iran}

\author{Reza Asgari}\email{asgari@ipm.ir}
\affiliation{School of Physics, Institute for Research in Fundamental Sciences (IPM), Tehran 19395-5531, Iran}

\begin{abstract}
We propose versatile materials based on the Ruderman-Kittel-Kasuya-Yosida (RKKY) interaction in a zigzag silicene nanoribbon (ZSNR)
on half filling in the presence of an out-of-plane electric field. We show that the topological phase transition in the band dispersion of ZSNR can be probed by using the RKKY interaction. We find that, due to the zero-energy edge states of the ZSNR, the exchange coupling is significantly enhanced when the impurities are located on the zigzag edges, and also explore that the strength of the interaction in the topological insulator phase is much greater than that when the system is in the band insulator region. We present a model to investigate the phase of a system of two magnetic impurities located on the edge of the ZSNR and find that three different magnetic phases, spiral, ferromagnetic, and anti-ferromagnetic, are possible for different values of the electric field. This electrical tunability of the magnetic phases in silicene can be explored by using current experimental techniques and can be of interest in the field of spintronics.

\end{abstract}

\maketitle

\section{Introduction}

The idea of spintronics is to utilize the spin degree of freedom of an electron associated with its intrinsic angular momentum \cite{MacNat,Yazyev2010}. This field is expected to play an important role in future technologies, such as information storage, processing and communication at faster speed and lower-energy consumption \cite{Power2013}. Among the different candidates in the field of spintronics including the $d$ and $f$ blocks of the periodic table, carbon and silicon have their own attractions, such as low density, biocompatibility, and plasticity. Although these atoms are not magnetic, there are some proposals to use their derivatives in the field of spintronics. One of the proposals is to use a doped system with magnetic adatoms.
Magnetic impurities in dilute magnetic semiconductors can interact indirectly via the itinerant electrons of the host material, which is known as the Ruderman-Kittel-Kasuya-Yosida (RKKY) interaction~\cite{Ruderman}. This interaction is a vital interaction and plays a significant role in the magnetic ordering of many electronic systems, including spin glasses~\cite{Egg} and alloys~\cite{Fu-sui}.

The RKKY interaction has been extensively studied in carbon based materials~\cite{Vozmediano,Dugaev,Brey,Bunder,Black-Schaffer,fariborz,MSashi1,MSashi2,Kogan,f2,Jelena} and it can be experimentally measured from single-atomic magnetometry of a pair of magnetic atoms \cite{Khajetoorians-2012,Zhou-2010} and also magnetotransport measurements \cite{Hindmarch-2003} based on the angle-resolved photoemission spectroscopy (ARPES) and scanning tunneling microscopy (STM) experiments. While this interaction has a simple Heisenberg term for spin nondegenerate materials such as graphene, it contains an additional Ising term in spin polarized graphene~\cite{fariborz}.
The Ising and Heisenberg terms of the RKKY coupling make the magnetic moment of impurities to be collinear. However, the Rashba or Dresselhaus terms lead to a Dzyaloshinskii-Moriya (DM) term, which causes a rotation between the spinors of the magnetic impurities. This phenomenon was shown to appear in three-dimensional topological insulators~\cite{Efimkin,Liu-Liu,zhu11} as well as a two-dimensional electron gas with the Rashba term~\cite{Imamura}. It has been also shown that the DM term can appear in a monolayer of molybdenum disulfide~\cite{Parhizgar mos2} and a graphene nanoribbon together with spin-orbit coupling~\cite{Jelena}, where the Rashba term is no longer present and the spin-valley coupling is the reason behind the twisted interaction between the impurities. In addition, the existence of the DM interaction has been reported for a bulk silicene~\cite{Xiao}.

Unlike carbon atoms in graphene, silicon atoms in silicene tend to adopt the tetrahedral $sp^3$ hybridization over $sp^2$, which results in a slightly buckled structure of silicene, where sublattices are displaced from each other in the out-of-plane direction with a distance {Takeda, Durgun, Cahangirov}. This buckled structure also leads to a relatively large spin-orbit coupling (SOC) ($\lambda_{SO}=3.9$ meV)~\cite{Liu} which is about 1000 times larger than graphene~\cite{Min, Yao}. The buckled honeycomb lattice and its Dirac band structure, combined with its sensitive surface, offers the potential for a widely tunable two-dimensional monolayer, where external fields can be exploited to influence the fundamental properties for future nanoelectronic devices~\cite{Tao, Tahir}. The buckled structure of silicene makes it possible to control its band dispersion by using an external out-of-plane electric field. In this scheme, the on-site energies of different sublattices are no longer equivalent and thus the spin-up and spin-down bands become separated~\cite{Ezawa, Drummond}.

A transition from a topological insulator state to a band insulator in silicene is predicted at $E_zd/2 =3.9$ meV by increasing the strength of the electric potential ~\cite{Ezawa,Kane-2005,Ezawa Nagaosa,cheng}.
Although there exists a finite gap in the bulk dispersion in both regions, the bulk-edge correspondence leads to a zero-gap structure in the topological phase, while such zero-energy edge states disappear in the band insulator state. Previously, the effect of this topological phase transition had been explored in silicene via Friedel oscillations related to the existence of a single impurity \cite{Chang-2014}.
In this work, we are aiming to find the effect of this topological phase transition in the nonlocal interaction between two magnetic impurities. We show that the topological phase can be probed using the RKKY interaction. To do so, we consider a finite-size ZSNRzigzag (see Fig.\ref{fig:rib}) and consider impurities to be located on the edge and calculate the RKKY interaction as a function of the distance between isilicene nanoribbon (ZSNR)mpurities for two different phases of normal and topological band insulators. To show that our findings are affected by the topological phase of the ZSNR, we compare our results with those when the impurities are located in the bulk. Moreover, to propose a practical application, we assume a system of two magnetic impurities on the edge of ZSNR and demonstrate that the ground state of such a system can displays spiral, ferromagnetic (FM), or antiferromagnetic (AFM) phases depending on the value of the out-of-plane electric field. Each of these phases has its own attraction in the field of spintronics~\cite{loktev}. Because of the electrical tunability of the magnetic phases of the system which is due to the electronic band tunability of silicene, it is of much interest to use magnetically doped silecene in nanotechnology~\cite{Ohno}.

The paper is organized as follows. In Sec.~\ref{sec:theory}, we first show the tight-binding Hamiltonian of a silicene nanoribbon in the presence of an out-of-plane electric field and then introduce the RKKY interaction in terms of noninteracting Green's function of the ZSNR. In Sec.~\ref{sec:numeric}, we present our numeric calculations for the coupling strengths of the RKKY interaction in the ZSNRs in terms of distance between the impurities. Also, different possible phases of magnetism for a system of two magnetic impurities on the edge of silicene are discovered. Section~\ref{sec:conc} contains discussions and a brief summary of our main results.

\section{The Model Hamiltonian and Formalism}\label{sec:theory}

The low-energy excitation of silicene is described by Dirac electrons with a large SOC interaction. The phase diagrams of silicene as functions of the electric potential and exchange field exhibit a rich physics~\cite{EzawaHall}, for instance, a quantum anomalous Hall insulator, valley polarized metal, quantum spin Hall, and band insulator phases appear. Remarkably, they are characterized by the Chern number and spin-Chern number together with the edge modes of a nanoribbon. The tight-binding Hamiltonian for a single layer silicene exposed to an external out-of-plane electric field $E_z$ can be written as~\cite{Liu,Ezawa}

\begin{eqnarray}\label{h}
\mathcal{H}&=&-t \sum_{\langle i,j\rangle s} {c_{is}^\dagger c_{js}}+i\frac{\lambda_{so}}{3\sqrt{3}} \sum_{\langle\langle i,j\rangle\rangle ss'} {\nu_{ij}c_{is}^\dagger \sigma_{ss'}^zc_{js'}}\nonumber\\
&+&l\sum_{is} {\zeta_i E_z^ic_{is}^\dagger c_{is}},
\end{eqnarray}
where $<>$ and $\ll\gg$ run over all the nearest and next-nearest neighbor hoping sites, respectively. $c_{i\sigma}^\dagger$ $(c_{i\sigma})$ creates (annihilates) an electron with spin $\sigma$ at site $i$, and $t=1.6$eV is the transfer energy. $\nu_{ij}={\bf d}_{ij}\cdot \hat{z}/|{\bf d}_{ij}|=\pm1$  $({\bf d}_{ij}={\bf d}_i\times {\bf d}_j)$ where ${\bf d}_{i},{\bf d}_{j}$ are the vectors from center of a hexagon to the $i$ and  $j$ adjacent to the same sublattices, respectively. $\zeta_i=\pm1$ for the $A$ or $B$ site, ${\bf \sigma}$ are the Pauli spin matrices and $l=d/2$. By performing a Fourier transformation in the continuum model, the low-energy effective Hamiltonian in the vicinity of the $K$ or $K'$ point is
\begin{eqnarray}\label{eq:H0k}
\mathcal{H}(k)=\hbar v_F( k_x \tau_x -  k_y\tau_y \eta)- \lambda_{SO} \tau_z \sigma_z \eta + lE_z\tau_z,
\end{eqnarray}
where $\eta= \pm 1$ for the $K$ or $K'$ Dirac point and ${\sigma_i}$,  ${\tau_i}$ are Pauli matrices for spin and sublattice pseudospin, respectively. The Fermi velocity is $v_{\rm F}=\frac{\sqrt{3}}{2\hbar}at$ with the lattice constant $a=3.86{\AA}$. The energy spectrum is thus given by

\begin{eqnarray}\label{eq:energy}
E_{\eta}(k)=\pm \sqrt{(\hbar v_F k)^2+(lE_z-s_z\eta\lambda_{SO})^2}.
\end{eqnarray}
In the absence of an external electric field, there are two separated bands with a SOC gap of $2\lambda_{so}$. However, by applying $E_z$, the inversion symmetry breaking causes the spin splitting of $|lE_z\pm\eta s_z\lambda_{so}|$ ~\cite{Yamakage}, where $s_z=\pm1$ denots the spin of electrons. The gap energy $E_g$ decreases linearly with increasing $E_z$ until a critical field $E_c=\eta s_z\lambda_{so}/l$.

\begin{figure}
\includegraphics[width=0.9\linewidth]{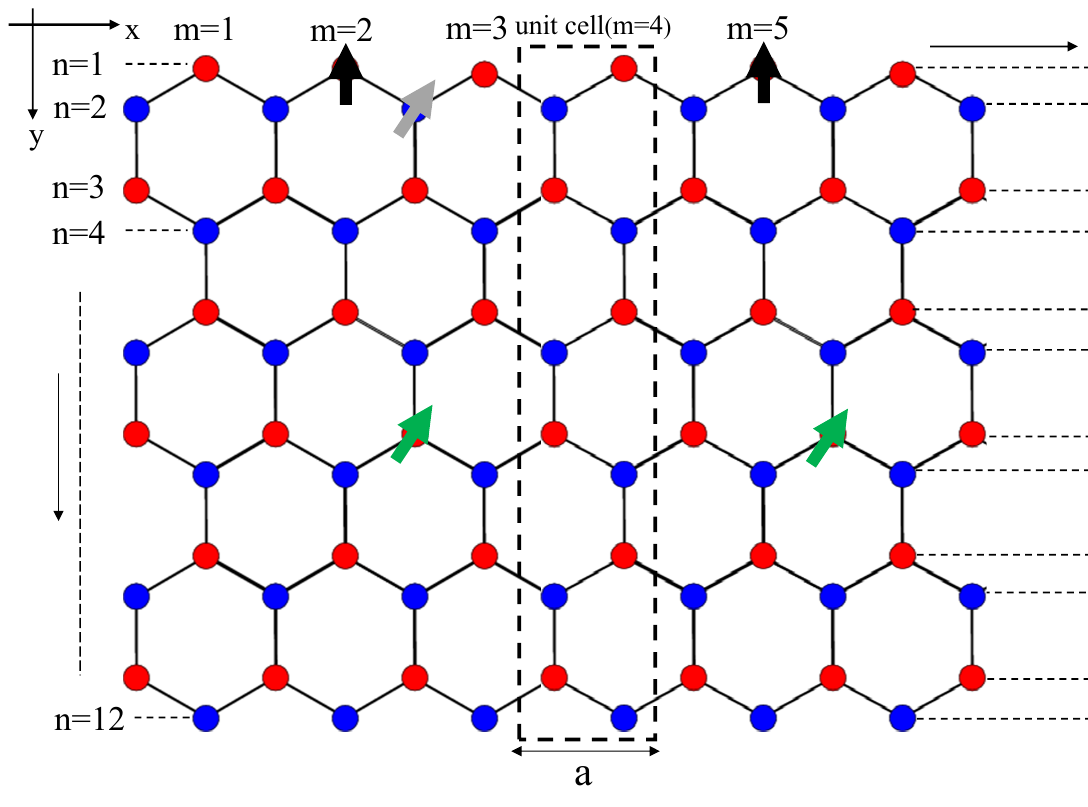}
\caption{(Color online) Lattice structure of the zigzag silicene nanoribbon. The edge and inside configurations are marked with black and green spins, respectively. The dashed box is a unit cell and $m,n$ represent the $x$ and $y$ coordinates of the lattice points. The length $(L)$ and the width $(N)$ of the nanoribbon are defined as the number of unit cells and the number of atoms in a unit cell, respectively. Here, we have taken $N=12$. 
\label{fig:rib}}
\end{figure}

Since it is possible to engineer the electronic structure of two-dimensional (2D) nanorribons by using scanning tunneling microscope lithography with nanometer precision of the patterning of nanoribbons~\cite{exp-nano}, we consider a zigzag silicene nanoribbon~\cite{Ding, An} with a finite width along the $y-$ direction, where its properties can be described by Eq.~(\ref{h}). A schematic representation of the ZSNR is shown in Fig.~\ref{fig:rib}. The dashed box is a unit cell and index $m$ represents the number of unit cells or equivalently the $x$ coordinate, and $n$ shows the $y$ coordinate of the lattice points, and therefore each site can be labeled by $(m,n)$, a notation that we use here to show the position of the impurity.

By locating magnetic impurities on the ZSNR, an exchange interaction between the spin of the magnetic impurities and the spin of
the conduction electrons of the silicene atoms occurs, which is given by
\begin{eqnarray}\label{eq:rkky}
{\mathcal{H}}_{int} = J_c \sum_{j=1,2} {\bf S(R_j)\cdot I_j},
\end{eqnarray}
where $J_c$ is the coupling constant between the conduction electrons and impurities, ${\bf I_j}$ is the magnetic moment of the $j$-th impurity at location ${\bf R_j}$, and ${\bf S(R)}=\frac{\hbar}{2} \sum_i \delta {\bf (r_i-R)} \sigma_i$, where $\sigma_i$ is the spin density operator of the $i-$conduction electron.

Using a second-order perturbation theory, one can replace the ${\mathcal{H}}_{int}$ by the RKKY interaction given by~\cite{Ruderman,Imamura}

\begin{eqnarray}\label{eq:Hrkky}
{{\mathcal{H}}^{\alpha\beta}_{RKKY}}= J_c^2 \sum_{k,l}{\bf I}_{1k}\chi^{kl}_{\alpha\beta}({\bf r,r'}){\bf I}_{2l}
\end{eqnarray}

Here, $\alpha$ and $\beta$ indicate the sublattices of which the impurities are located on and $\chi_{kl}$ ($k,l=x,y,$ and $z$) is the spin susceptibility tensor of the system in real space, obtained through the retarded Green's functions $G_{\alpha\beta}^{0}$,
\begin{eqnarray}
\chi_{\alpha\beta}^{k,l}({\bf r,r'})=-\frac{2}{\pi}\Im m\int_{-\infty}^{\varepsilon_{\rm f}}d\varepsilon Tr[ \sigma^k G_{\alpha\beta}^{0}({\bf r,r'},\varepsilon)\sigma^l G_{\beta\alpha}^{0}({\bf r',r},\varepsilon)],\nonumber\\
\end{eqnarray}
where $\varepsilon_{\rm F}$ is the Fermi energy, and the trace is over the spin degree of freedom.
                                                 the
In order to calculate the spin susceptibility of the finite-size silicene nanoribbon, we use the spectral representation of the noninteracting Green's function~\cite{Sherafatiribbon},
\begin{eqnarray}\label{eq:Green}
G_{\alpha\beta}^{0}({\bf r,r'},\varepsilon)=\sum_{n,s}\frac{{\psi_{n,s}^{\alpha}}({\bf r}){\psi_{n,s}^{\ast\beta}}(\bf r')}{\varepsilon-\varepsilon_{n,s}+i\eta},
\end{eqnarray}
where the sum runs over all eigenstates $\psi_{n,s}$ and corresponding eigenvalues $\varepsilon_{n,s}$ with spin $s$ and the band index $n$ obtained by diagonalizing the finite-size Hamiltonian, Eq.~(\ref{h}).
The RKKY Hamiltonian, can be written as
\begin{eqnarray}
\mathcal{H}_{RKKY} &=&J_H {\bf I}_{1}\cdot{\bf I}_{2}+J_{DM}({\bf I}_1 \times {\bf I}_2)\cdot\hat{z}
+J_{I}I_{1z}I_{2z},
\end{eqnarray}
where $J_H=J^2_c\chi_{xx}$, $J_{DM}=J^2_c \chi_{xy}$ and $J_I=J^2_c(\chi_{zz}-\chi_{xx})$ represents the Heisenberg, Dzyalosinskii-Moriya, and Ising terms, respectively, and all the other parts of the spin susceptibility are zero.
The explicit form of $\chi_{ij}$ as a function of eigenvectors and eigenvalues of the system is given in the Appendix.

\section{Numerical results}\label{sec:numeric}

The general features of the exchange coupling, basically the dependence of the RKKY interaction on the distance $R$, have been numerically studied previously~\cite{Xiao} for a bulk silicene system.
Here, we mainly focus on the half filling $(\varepsilon_F=0)$ for various configurations for which the positions of two magnetic impurities are either inside the ZSNR or on the edges. Our main purpose here is to explore the topological phase and edge states dependence of the RKKY interaction. In this regard, we consider two electric field strengths in which their corresponding bulk gaps are equal. For instance, we choose $l E_z=1$ meV, and $l E_z=6.8$ meV, where the bulk band gaps of these two fields are both $2.9$ meV, and then calculate the RKKY interaction between the two magnetic impurities as a function of distance between them for different impurity configurations.

\begin{figure}
\includegraphics[width=1.0\linewidth]{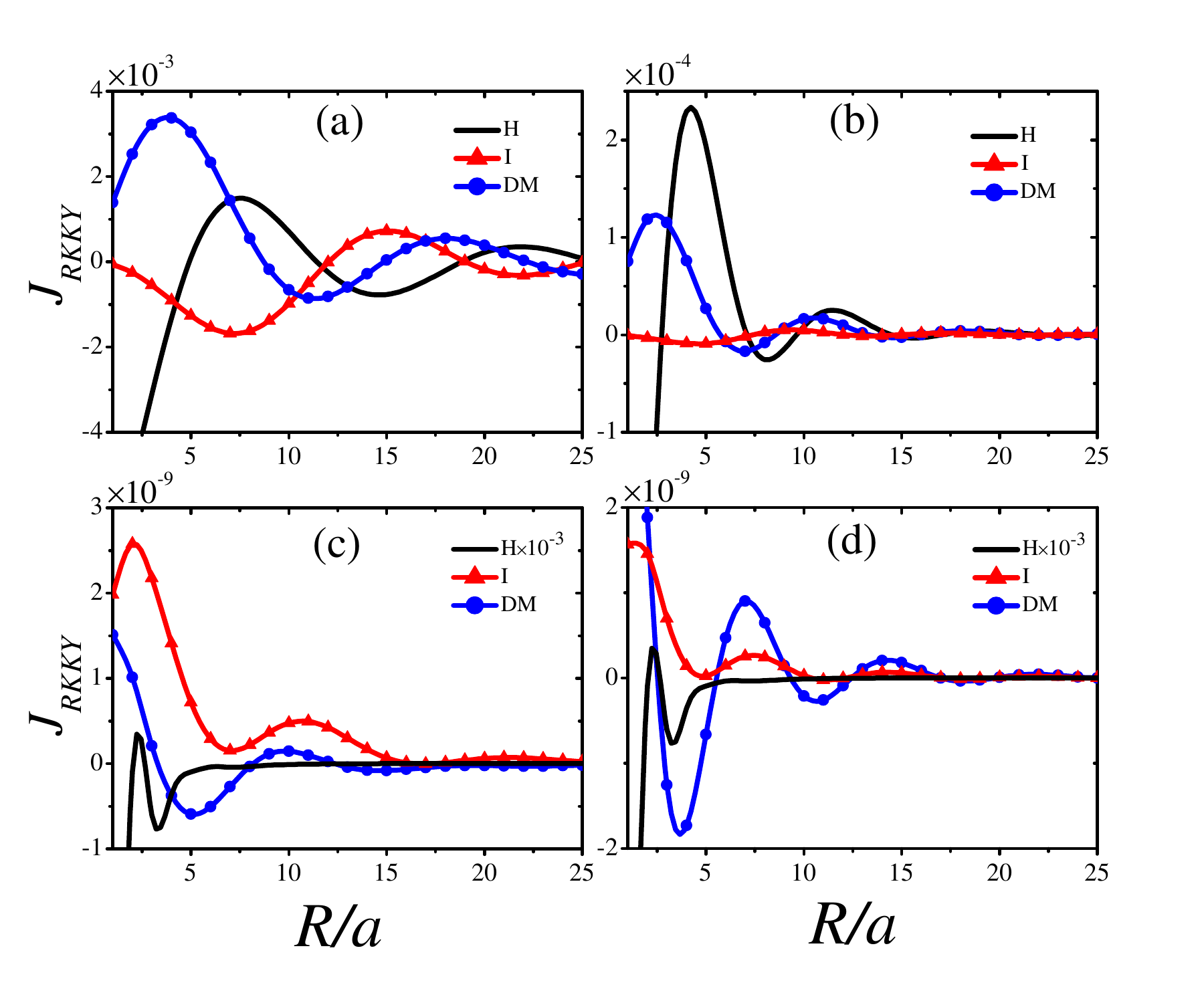}
\caption{(Color online) Various terms of the RKKY interaction scaled by $J^2_c $ as a function of the impurity distances for the ZSNR with $N=12, L=300$.(a),(b) Impurities located on the edge(black arrows) such that the first impurity is fixed at the sublattice $(11,1)$ and the second one is located at (m,1) where $m=12, 13,...$. (c),( d) Impurities is located inside the ZSNR(green arrows) such that the first impurity located on the sublattice $(11,6)$ and the second one is placed at $(m,6)$ with $m=12, 13,...$. In these figures, the electric field is fixed at $lE_z=1$ meV in (a) and (c) $lE_z=6.8$ meV in (b) and (d). Note that $J_{RKKY}$ shows an oscillatory behavior in $R$ and decays fast with a short-ranged behavior. The RKKY coupling is significantly larger than the situation where both impurities are in the bulk due to the existence of nearly-zero-energy states at the edge of the ZSNR.
\label{fig:ed}}
\end{figure}

Figures ~\ref{fig:ed} (a) and ~\ref{fig:ed} (b) show the spatial behavior of the RKKY interaction for two impurities, both sitting on the same edge along the zigzag direction for $N=12$ and $L=300$ Fig.~\ref{fig:ed} (a)$lE_z=1$ meV and Fig.~\ref{fig:ed} (b)$lE_z=6.8$ meV. Figures ~\ref{fig:ed} (c) and ~\ref{fig:ed} (d) are the same as Figs.~\ref{fig:ed} (a) and ~\ref{fig:ed} (b), but for two impurities located inside the zigzag nanoribbon along the line $n=6$. As shown in this figure, $J_{RKKY}$ for all terms displays an oscillatory behavior with respect to the distance between two impurities. As an important result, by comparing Figs.~\ref{fig:ed} (a) and ~\ref{fig:ed} (c) and Figs.~\ref{fig:ed} (b) and ~\ref{fig:ed} (d), the RKKY coupling in the cases shown in Figs.~\ref{fig:ed} (a) and ~\ref{fig:ed} (b) is significantly larger than when both impurities are in the bulk due to the existence of nearly-zero-energy states at the edge of the ZSNR ~\cite{Ezawa Nagaosa}. This fact is true even in the cases where the system is in the band insulator regime [Figs.~\ref{fig:ed} (b) and ~\ref{fig:ed} (d)] since the local density of states drops dramatically by going from the edge to the bulk~\cite{hsin}, and therefore the RKKY interaction is much larger in the case shown in Fig.~\ref{fig:ed} (b) in comparison with results demonstrated in Fig.~\ref{fig:ed} (d). In addition, in the topological insulator region ($lE_z=1$ meV$<\lambda_{so}$), the RKKY interaction is about 20 times greater than in the band insulator region ($lE_z=6.8$ meV$>\lambda_{so}$) when impurities are located on the edge, as shown in Figs.~\ref{fig:ed} (a) and ~\ref{fig:ed} (b), while no difference in the order of magnitude  appears in the case where impurities are in the bulk, as shown in Figs.~\ref{fig:ed} (c) and ~\ref{fig:ed} (d). Note that although $J_{RKKY}$ shows a few oscillations in $R$ owing to the smallness of the gap of silicene, it decays fast with a short-ranged behavior. We also calculate the RKKY strength in the case of one impurity on the edge $(11,1)$ and the other one on the inner edge on the line $(m,2)$ with $m=12, 13,...$ (gray arrow in Fig. \ref{fig:rib}). The obtained data are qualitatively similar to those results shown in Figs.~\ref{fig:ed} (c) and ~\ref{fig:ed} (d) which explore that both impurities might be sitting on the edge to obtain the topological effect.

\begin{figure}
\includegraphics[width=0.9\linewidth]{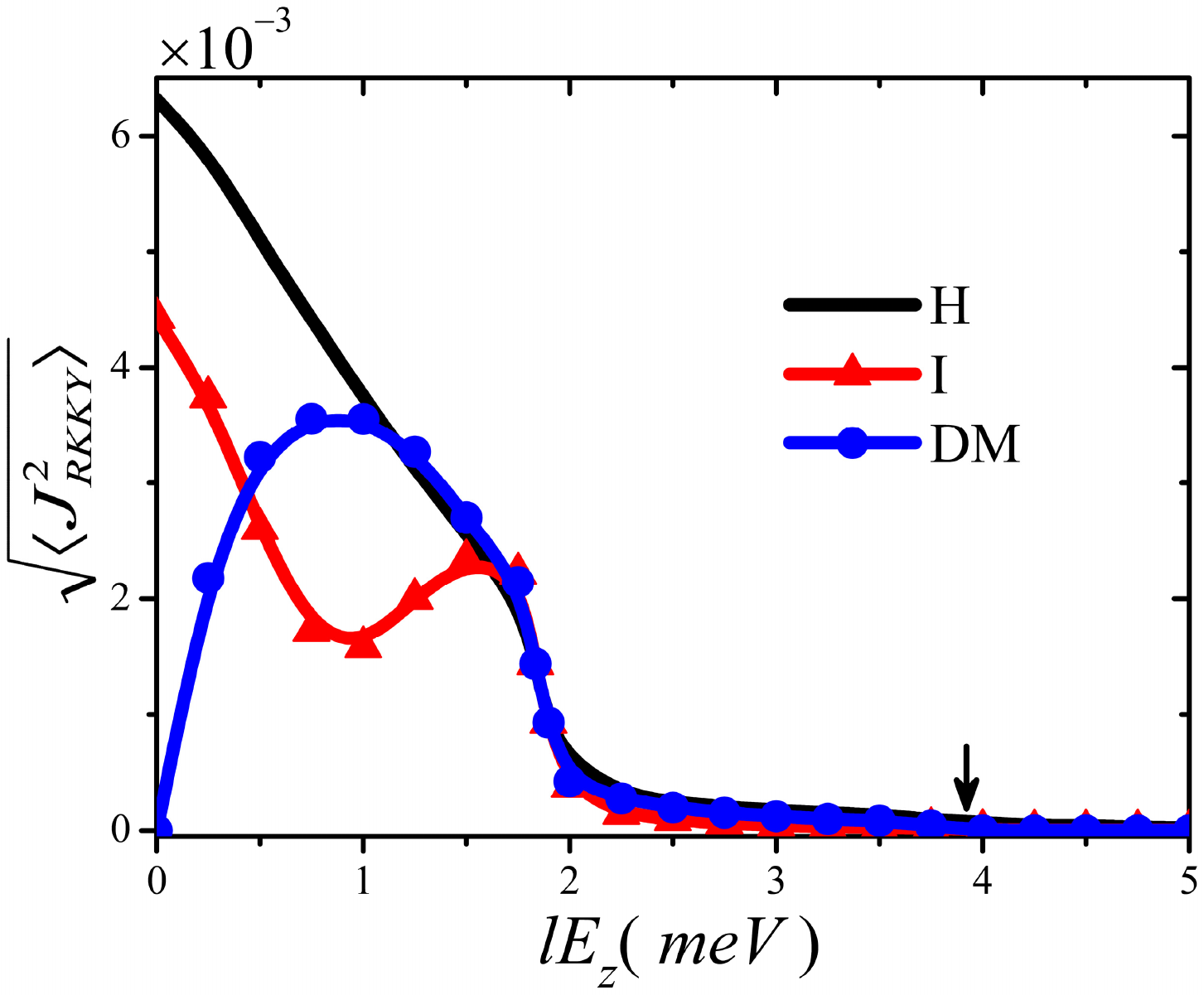}
\caption{(Color online) Root mean square of various terms of the RKKY interaction scaled by $J^2_c$ as a function of the perpendicular electric potential for the ZSNR with $N=22,L=300$. Here, the first impurity is fixed at the sublattice $(147,1)$ and the position of the second one is varied from sublattices $(148,1)$ to $(157,1)$. $E_c$ in this figure has been labeled by a black vertical arrow.
\label{fig:ez1}}
\end{figure}

We calculate the electric field dependence of the RKKY interaction for different distances between magnetic impurities located on the edge to explore the effect of the topological phase transition on the RKKY interaction. Figure ~\ref{fig:ez1} shows the electric potential dependence of the various terms of the RKKY coupling, when both impurities are located on the edge of ZSNR. To include the effect of different distances between impurities, we consider impurities with distances from $1a$ to $10a$ on the edge and take the root mean square value of the RKKY interaction. As seen in Fig.~\ref{fig:ez1}, in the topological insulator region ($E_z<E_c$), all the interaction coefficients are much greater than those results obtained for the band insulator region ($E_z>E_c$). Note that this behavior is special to the edge of ZSNR and as one calculates the same for the bulk no change in the order of magnitude can be seen for different phases.
\begin{figure}
\includegraphics[width=0.9\linewidth]{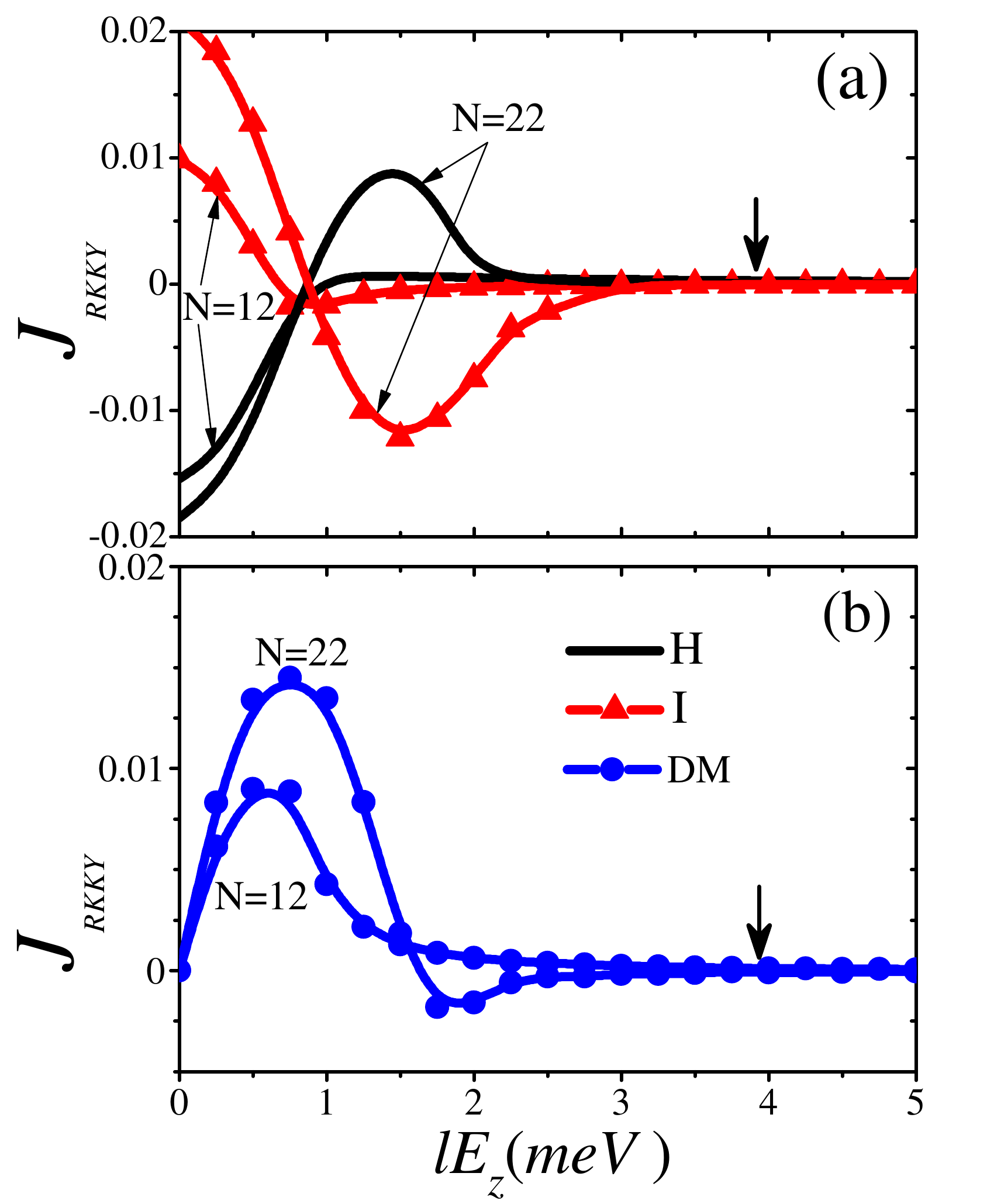}
\caption{(Color online) Various terms of the RKKY interaction scaled by $J^2_c $ as a function of the perpendicular electric potential for $L=300$, when two impurities are fixed at the zigzag edge at $(148,1)$ and $(153,1)$ lattice points (a) the Heisenberg and Ising terms and (b) the Dzyaloshinsky-Moriya term. In both cases $R/a=5$. Noticeably, the RKKY coupling increases with ZSNR width due to the increase of the density of states of the edge mode. Black vertical arrows show the position of $E_c$.
\label{fig:ez4}}
\end{figure}

We compare various terms of the RKKY interaction for two widths of the ZSNR in Fig.~\ref{fig:ez4} to study the effect of the width of ZSNR on the RKKY coupling. We show the RKKY interaction (scaled by $J^2_c $) as a function of the perpendicular electric potential, for two impurities located with a fixed distance on the zigzag edge at $(148,1)$ and $(153,1)$ lattice points. Figure ~\ref{fig:ez4} (a) shows the Heisenberg and the Ising terms and Fig.~\ref{fig:ez4} (a) belongs to the Dzyaloshinskii-Moriya term. In both cases $R/a=5$. By comparing the results of various systems with different widths, it is clearly found that the RKKY coupling increases with the ZSNR width, as reported previously~\cite{Ezawa Nagaosa,wakabayashi}. This means that thicker nanoribbons can show better topological effects in ZSNRs.
Although we assume a narrow nanoribbon in our numerical calculations, we expect that sharper differences between the results of the bulk and edge states should be appear in the thicker cases too.
In contrast to changing the width of ZSNR, we find that changing the length of the sample has little effect on the RKKY interaction.
Note that since we assume the Fermi energy to be zero in these figures, all the diagrams are short ranged and fall off rapidly with an exponential like function.
Most importantly, we examine the RKKY interaction for an n-doped ZSNR and we find that the RKKY coupling behaves as $R^{-1}$ in the finite values of the doping which shows that the edges of doped ZSNR are acting as a one-dimensional (1D) semiconductor.

Now, we assume a system of two magnetic impurities located on the edge of ZSNR with constant distance. This would be a useful and practical application of the RKKY interaction in a nanoribbon of silicene. Our aim is to explore the final phase of such a system. Following Ref. ~\cite{Parhizgar mos2}, we assume magnetic impurities as classical spin vectors and rewrite the RKKY interaction Hamiltonian in the spherical coordinate. We find that the equilibrium direction of the vectors of the magnetic moments is $\theta_{1,2}=0$, $\pi$ or $\theta_{1}=\theta_{2}=\pi/2$, where $\theta$ is the polar angle and it is defined as an angle between the spin direction with the unit vector normal to the ZSNR surface ($\hat{z})$. Also, we find that the impurity moments shall be located inside the plane with a relative azimuthal angle of $\phi=\tan^{-1}(J_{DM}/J_H)$, if two quantities $D_1=-sign(J_H)\sqrt{J_H^2+J_{DM}^2}$, $D_2=J_{DM}^2-J_I(2J_H+J_I)$ are both positive ($D_1$ and $D_2$ are quantities of the Hessian matrix introduced in Ref~[\onlinecite{Parhizgar mos2})]. Otherwise, the magnetic moment of impurities will be perpendicular to the plane and will form a FM structure when $J_H+J_I<0$ or an AFM structure for the case when $J_H+J_I>0$.

\begin{figure}
\includegraphics[width=0.95\linewidth]{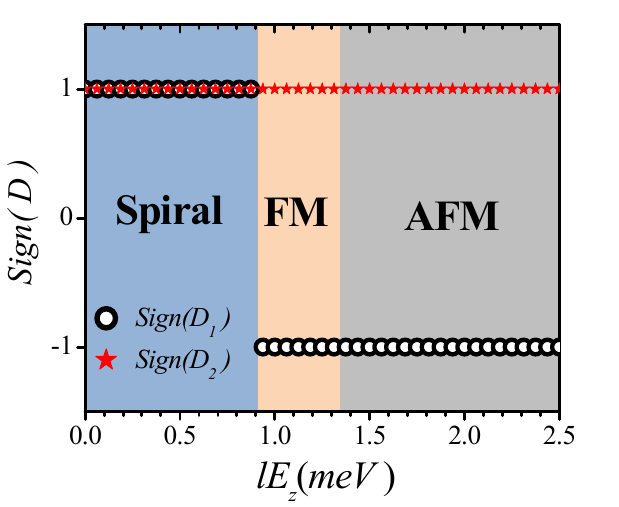}
\caption{(Color online) Sign ($D_{1,2}$) and phase diagram of a system of two magnetic impurities placed on the edge of ZSNR with $N=12,L=300$ as a function of the external potential.
Here, impurities are located with five lattice vectors ($R/a=5$) from each other, at $(148,1)$ and $(153,1)$ lattice points.
\label{fig:sign}}
\end{figure}

Figure~\ref{fig:sign} shows the sign of $D_1$ and $D_2$ as a function of the out-of-plane electric potential $lE_z$ at half filling in the case where impurities are located on the edge of the nanoribbon with distance $R=5a$. A change of the equilibrium state of the system from the case where impurities have a finite angle to the case where they have a parallel, FM-like structure, occurs at $lE_z\sim 0.91$ meV. Afterwards, another equilibrium state occurs from parallel to antiparallel magnetic moments at $lE_z\sim 1.35$ meV.
It would be worthwhile mentioning that in the spiral region, the direction of the magnetic moment is in the plane of the sample such that each impurity rotates with an angle $\phi$ with respect to the next one. This angle of rotation starts from zero and increases up to high values of about $80\AA$ as $E_z$ increases.

\section{conclusion}\label{sec:conc}

In conclusion, we have studied the effect of the out-of-plane electric field and topological phase transition on the RKKY interaction in an undoped silicene nanoribbon.
The band dispersion of silicene is sensitive to the applied electric field and posseses a topological phase transition from a topological insulator to a band insulator. Using the lattice Green's function technique in Lehmann's representation we have derived different terms of spin susceptibility, which is related to the RKKY interaction. The RKKY interaction will have three different terms, namely, Heisenberg, Ising, and Dzyalosinkii-Moriya. As a result of the bulk-edge correspondence in the topological insulator phase, the gap closes at the boundary of the nanoribbon, and this shows its effect on the RKKY interaction for the impurities located on the edge. We have shown that, although the magnetic coupling decays fast and is short ranged, the RKKY interaction is significantly large in the case where two magnetic moments are located on the nanoribbon edges rather than sitting in the sheets. Remarkably, in the topological regime, the RKKY interaction is about 20 times greater than in the bulk insulator regime.

One of the fascinating features in this case is the possibility for the enhancement of the interaction by tuning the electric field, which opens an avenue to probe the interaction experimentally. In addition, we have demonstrated that the ground state of the system consisting of dilute magnetic moments on the edge can display a spiral, ferromagnetic, or antiferromagnetic phase depending on the value of the out-of-plane electric field. This electrical tunability of the magnetic phases in  dilute magnetic silicene can be investigated using current experimental techniques and can be of interest in the field of spintronics.

As the final point, we want to address the experimental accessibility of our proposals.
The RKKY interaction is small such that measuring it experimentally is usually hard. However, using a high value of density of states on the edge of materials such as zigzag graphene nanoribbons \cite{Black-Schaffer}, or band tunability of materials such as bilayer graphene \cite{f2}, increases the chance to measure this quantity. Zero gapped edge states in graphene only happen in perfect zigzag edges,
however, there are both edge modes and band tunability in silicence.
We should mention that, although we carried out the calculation for a zigzag silicene nanoribbon, since silicene is a Kane-Mele topological insulator\cite{Kane-2005}, the edge modes are preserved even for rough edges. This fact increases the possibility that our theoretical predictions may be explored in experiment.

\appendix
\section{Explicit form of the spin susceptibility of a nanoribbon}\label{sec:app}
In this section, we want to show the form of the spin susceptibility as a function of eigenvalues and eigenvectors of a ZSNR.
 Since we neglect the Rashba term, the Green's function for spin-up and spin-down can be separated in the matrix form $G=
\begin{pmatrix}
g^{(\prime)}_{\uparrow}&0\\
0&g^{(\prime)}_\downarrow\\
\end{pmatrix}
 $
where $g_{\uparrow(\downarrow)}$ and $g'_{\uparrow(\downarrow)}$ are the spin components of $G({\bf r,r'},\varepsilon)$ and $G({\bf r',r},\varepsilon)$, respectively.
The various components of the spin susceptibility are given by $\chi^{k,l}({\bf r,r'})=-\frac{2}{\pi}\Im m\int_{-\infty}^{\varepsilon_F}d\varepsilon \nu^{k,l}({\bf r,r'})$, where $\nu^{k,l}({\bf r,r'})= Tr[\sigma^k G^{0}({\bf r},{\bf r'},\varepsilon)\sigma^l G^{0}({\bf r',r},\varepsilon)]$ and is given by
\begin{eqnarray}\label{eq:nu}
\nu^{xx}&=& \nu^{yy}=g_\downarrow g'_\uparrow+g_\uparrow g'_\downarrow\nonumber\\
\nu^{zz}&=&g_\uparrow g'_\uparrow+g_\downarrow g'_\downarrow\nonumber\\
\nu^{xy}&=&-i(g_\uparrow g'_\downarrow-g_\downarrow g'_\uparrow)
\end{eqnarray}
while the rest of the terms are zero.

In order to calculate the spin susceptibility, we apply the Lehmann's representation of the Green's function~\eqref{eq:Green}. Each term of the multiplication of two Green functions can be rewritten in a compact form as $g_s g'_{s'}=\sum_{n,n'}\Psi^{ss'}_{nn'}E^{ss'}_{nn'}$ where $\Psi^{ss'}_{nn'}=\psi_{n}^{s}({\bf r})\psi_{n}^{\ast s}({\bf r'})\psi_{n'}^{s'}({\bf r'})\psi_{n'}^{\ast s'}({\bf r})$ and
$E_{nn'}^{ss'}=[({\varepsilon-\varepsilon_{n}^{s}+i\eta})({\varepsilon-\varepsilon_{n'}^{s'}+i\eta})]^{-1}$.
Note that under the interchange of $n$ and $n'$,  $g_{s} g'_{s'}\longrightarrow\sum_{n,n'}\Psi^{\ast s's}_{nn'}E^{s's}_{nn'}$ and in a similar manner $g_{s'} g'_{s}$ converts to  $\sum_{n,n'}\Psi^{\ast ss'}_{nn'}E^{ss'}_{nn'}$.

By applying these properties to the Eq.~\eqref{eq:nu} and using $\lim_{\eta\to 0^+} (x\pm i\eta)^{-1}=P(1/x)\mp i\pi \delta(x)$, and some straight forward calculations one can show
\begin{eqnarray}
\chi^{xx}({\bf r,r'})&=&2\sum_{n\neq n'}\frac{f(\varepsilon_n^{\uparrow})-f(\varepsilon_{n'}^{\downarrow})}{\varepsilon_n^{\uparrow}-\varepsilon_{n'}^{\downarrow}}\Re e [\psi_n^{\uparrow}({\bf r})\psi_n^{\ast\uparrow}({\bf r}')\psi_{n'}^{\downarrow}({\bf r}')\psi_{n'}^{\ast\downarrow}({\bf r})] \nonumber\\
&+&\frac{f(\varepsilon_n^{\downarrow})-f(\varepsilon_{n'}^{\uparrow})}{\varepsilon_n^{\downarrow}-\varepsilon_{n'}^{\uparrow}}\Re e[\psi_n^{\downarrow}({\bf r})\psi_n^{\ast\downarrow}({\bf r}')\psi_{n'}^{\uparrow}({\bf r}')\psi_{n'}^{\ast\uparrow}({\bf r})].
\end{eqnarray}

In analogy to the $\chi^{xx}$, the $\chi^{zz}$ yields
\begin{eqnarray}
\chi^{zz}({\bf r,r'})&=&2\sum_{n\neq n'}\frac{f(\varepsilon_n^{\uparrow})-f(\varepsilon_{n'}^{\uparrow})}{\varepsilon_n^{\uparrow}-\varepsilon_{n'}^{\uparrow}}\Re e[\psi_n^{\uparrow}({\bf r})\psi_n^{\ast\uparrow}({\bf r}')\psi_{n'}^{\uparrow}({\bf r}')\psi_{n'}^{\ast\uparrow}({\bf r})] \nonumber\\
&+&\frac{f(\varepsilon_n^{\downarrow})-f(\varepsilon_{n'}^{\downarrow})}{\varepsilon_n^{\downarrow}-\varepsilon_{n'}^{\downarrow}}\Re e[\psi_n^{\downarrow}({\bf r})\psi_n^{\ast\downarrow}({\bf r}')\psi_{n'}^{\downarrow}({\bf r}')\psi_{n'}^{\ast\downarrow}({\bf r})],
\end{eqnarray}
and finally in a similar manner, $\chi^{xy}$ can be written as
\begin{eqnarray}
\chi^{xy}({\bf r,r'}) &=& 2\sum_{n\neq n'}\frac{f(\varepsilon_n^{\uparrow})-f(\varepsilon_{n'}^{\downarrow})}{\varepsilon_n^{\uparrow}-\varepsilon_{n'}^{\downarrow}}\Im m[\psi_n^{\uparrow}({\bf r})\psi_n^{\ast\uparrow}({\bf r}')\psi_{n'}^{\downarrow}({\bf r}')\psi_{n'}^{\ast\downarrow}({\bf r})]\nonumber\\ &-&\frac{f(\varepsilon_n^{\downarrow})-f(\varepsilon_{n'}^{\uparrow})}{\varepsilon_n^{\downarrow}-\varepsilon_{n'}^{\uparrow}}\Im m[\psi_n^{\downarrow}({\bf r})\psi_n^{\ast\downarrow}({\bf r}')\psi_{n'}^{\uparrow}({\bf r}')\psi_{n'}^{\ast\uparrow}({\bf r})].\nonumber\\
\end{eqnarray}



\end{document}